\documentclass[12pt,letterpaper]{article}
\usepackage[margin=1in]{geometry}
\usepackage[english]{babel}
\usepackage[utf8x]{inputenc}
\usepackage[T1]{fontenc}

\usepackage{amsmath}
\usepackage{amssymb}
\usepackage{graphicx}
\usepackage{adjustbox}
\usepackage[colorinlistoftodos]{todonotes}
\usepackage[colorlinks=true, allcolors=blue]{hyperref}
\usepackage{subfigure}
\usepackage[toc,page]{appendix} 
\usepackage{wrapfig}
\usepackage{natbib}
\usepackage{aas_macros}
\usepackage{pifont}
\newcommand{\cmark}{\ding{51}}%

\newcommand{\done}{\rlap{$\square$}{\raisebox{2pt}{\large\hspace{1pt}\cmark}}}%

\begin{document}
\date{}

\title{\bf Astro2020 Science White Paper\\\vspace{0.2in} Probing Feedback in Galaxy Formation \\ with Millimeter-wave Observations}
\maketitle

\noindent \textbf{Thematic Areas:} \hspace*{60pt} $\square$ Planetary Systems \hspace*{10pt} $\square$ Star and Planet Formation \hspace*{20pt}\linebreak
$\square$ Formation and Evolution of Compact Objects \hspace*{20pt} $\done$ Cosmology and Fundamental Physics \linebreak
  $\square$  Stars and Stellar Evolution \hspace*{1pt} $\square$ Resolved Stellar Populations and their Environments \hspace*{20pt} \linebreak
  $\done$    Galaxy Evolution   \hspace*{45pt} $\square$             Multi-Messenger Astronomy and Astrophysics \hspace*{65pt} \linebreak

\noindent\textbf{Principal Authors:}
\\
Names: Nicholas Battaglia, J.~Colin Hill\\
Institutions: Cornell University, Institute for Advanced Study\\
Emails: \texttt{nbatta@astro.cornell.edu}, \texttt{jch@ias.edu}\\
Phones: (607)-255-3735, (509)-220-8589\\

\noindent\textbf{Co-authors:}
  \linebreak
Stefania Amodeo (Cornell), James G.~Bartlett (APC/U.~Paris Diderot), Kaustuv Basu (University of Bonn), Jens Erler (University of Bonn), Simone Ferraro (Lawrence Berkeley National Laboratory), Lars Hernquist (Harvard), Mathew Madhavacheril (Princeton), Matthew McQuinn (University of Washington), Tony Mroczkowski (European Southern Observatory), Daisuke Nagai (Yale), Emmanuel Schaan (Lawrence Berkeley National Laboratory), Rachel Somerville (Rutgers/Flatiron Institute), Rashid Sunyaev (MPA), Mark Vogelsberger (MIT), Jessica Werk (University of Washington)
\\

\noindent\textbf{Endorsers:}
  \linebreak
James Aguirre (University of Pennsylvania), Zeeshan Ahmed (SLAC), Marcelo Alvarez (UC--Berkeley), Daniel Angles-Alcazar (Flatiron Institute),  Chetan Bavdhankar (National Center for Nuclear Physics), Eric Baxter (University of Pennsylvania), Andrew Benson (Carnegie Observatories), Bradford Benson (Fermi National Accelerator Laboratory/ University of Chicago), Paolo de Bernardis (Sapienza Universit\`{a} di Roma), Frank Bertoldi (University of Bonn), Fredirico Bianchini (University of Melbourne), Colin Bischoff (University of Cincinnati), Lindsey Bleem (Argonne National Laboratory/KICP), J.~Richard Bond (University of Toronto/ CITA), Greg Bryan (Columbia/Flatiron Institute), Erminia Calabrese (Cardiff), John E.~Carlstrom (University of Chicago/Argonne National Laboratory/KICP), Joanne D.~Cohn (UC--Berkeley), Asantha Cooray (UC--Irvine), William Coulton (Cambridge), Thomas Crawford (University of Chicago/KICP), Marco De Petris (Sapienza Universit\`{a} di Roma), Kelly A.~Douglass (University of Rochester), Joanna Dunkley (Princeton), Alexander van Engelen (CITA), Giulio Fabbian (University of Sussex), Simon Foreman (CITA), Shaul Hanany (University of Minnesota), Christopher Hayward (Flatiron Institute), Terry Herter (Cornell), Christopher M. Hirata (Ohio State University), Philip F.~Hopkins (Caltech), Bhuvnesh Jain (University of Pennsylvania), Bradley Johnson (Columbia), Theodore Kisner (Lawrence Berkeley National Laboratory), Lloyd Knox (UC--Davis), Ely D.~Kovetz (Ben-Gurion University), Andrey Kravtsov (University of Chicago), Benjamin L'Huillier (KASI), Massimiliano Lattanzi (INFNFE), Adam Lidz (University of Pennsylvania), Adam Mantz (Stanford), Kiyoshi Masui (MIT), Jeff McMahon (University of Michigan), Joel Meyers (SMU), Suvodip Mukherjee (IAP), Julian B.~Mu\~noz (Harvard),  Michael Niemack (Cornell), Andrei Nomerotski (Brookhaven National Laboratory), Lyman Page (Princeton), Bruce Partridge (Haverford), Levon Pogosian (Simon Fraser University), Clement Pryke (University of Minnesota), Giuseppe Puglisi (Stanford, KIPAC), Eliot Quataert (UC--Berkeley), Mathieu Remazeilles (University of Manchester), Douglas Scott (University of British Columbia), Blake D.~Sherwin (Cambridge/DAMTP), Sara Simon (University of Michigan), An\v{z}e Slosar (Brookhaven National Laboratory), David Spergel (Princeton/Flatiron Institute), Suzanne Staggs (Princeton), George Stein (University of Toronto), Radek Stompor (APC), Aritoki Suzuki (Lawrence Berkeley National Laboratory), Stephanie Tonnesen (Flatiron Institute), Yu-Dai Tsai (Fermi National Accelerator Laboratory), Caterina Umilt\`a (University of Cincinnati), Abigail Vieregg (University of Chicago), Edward J.~Wollack (NASA Goddard Space Flight Center), W.~L.~K.~Wu (KICP), Siavash Yasini (University of Southern California), Ningfeng Zhu (University of Pennsylvania)

 \begin{abstract}

Achieving a precise understanding of galaxy formation in a cosmological context is one of the great challenges in theoretical astrophysics, due to the vast range of spatial scales involved in the relevant physical processes.  Observations in the millimeter bands, particularly those using the cosmic microwave background (CMB) radiation as a ``backlight'', provide a unique probe of the thermodynamics of these processes, with the capability to directly measure the density, pressure, and temperature of ionized gas.  Moreover, these observations have uniquely high sensitivity into the outskirts of the halos of galaxies and clusters, including systems at high redshift.  In the next decade, the combination of large spectroscopic and photometric optical galaxy surveys and wide-field, low-noise CMB surveys will transform our understanding of galaxy formation via these probes.
 
\end{abstract}   

\newpage \setcounter{page}{1}

\section{Motivation}
\vspace{-2mm}

The past decade has seen tremendous advances in our theoretical and computational understanding of how galaxies form and evolve over cosmic time.  To a large extent, these breakthroughs have resulted from dedicated efforts to observe large samples of galaxies across the electromagnetic spectrum via wide-field surveys.  On the theoretical front, improvements in computing infrastructure and numerical methods have enabled extraordinary progress in simulating galaxy evolution in cosmological volumes.  These efforts have culminated in recent magneto-hydrodynamical
simulations in boxes that span hundreds of Mpc, which reproduce many multi-wavelength properties of galaxies, particularly their stellar populations~\citep[e.g.,][]{Pillepich2018}. 

{\bf The next major goal for theoretical and computational galaxy formation is understanding the physical processes and thermodynamic properties that govern the ionized baryons in galaxies and clusters.}  The vast majority of baryons in these systems are not contained in stars, but rather in the circumgalactic medium (CGM) and the intracluster medium (ICM).\footnote{Throughout, we refer to galaxy groups and clusters as ``clusters'', and their ionized gas as the ``ICM''.}  The CGM and ICM are the baryonic reservoirs that govern the evolution of galaxies and clusters, the luminous peaks in the large-scale structure of the Universe.  Encoded in their thermodynamic properties are the effects of assembly history and feedback processes that shape galaxy and cluster formation.  In particular, it is now clear that a precise understanding of the energetics of feedback in the CGM and ICM is required to narrow down the space of structure formation models~\citep[e.g.,][]{vandeVoort2016,Hillinprep}.

Multiple feedback processes affect the CGM and ICM.  The current standard view is that in galaxies with masses close to or below that of the Milky Way, feedback processes due to stellar winds and supernovae (hereafter referred to as stellar feedback) are most important~\citep[e.g.,][]{RS2015}.  For more massive galaxies, and in groups and clusters, it is thought that feedback from active galactic nuclei (AGN) is the dominant mechanism that regulates star formation.  However, since clusters form hierarchically, it is likely that stellar feedback processes were important at earlier times and played a critical role in shaping the thermodynamic properties of the progenitor halos that later merged to form these massive systems.  {\bf Probing the thermodynamic properties of the CGM and ICM across wide ranges in mass and radial scale, and especially out to higher redshifts near the peak of cosmic star formation, will provide critical information on feedback mechanisms.}

Large cosmological simulations that include baryons provide a partially predictive model for galaxy formation, tracking the evolution of gas and stars inside and outside of galaxies.  These simulations are now able to reproduce the optical properties of galaxies measured by the Hubble Space Telescope and ground-based optical surveys, such as the luminosity function, the stellar mass function, and the fractions of red and blue galaxies \citep[e.g.,][]{fire,Eagles,HorizonAGN,tng2018}. However, these simulations do not resolve the scales necessary to perform {\it ab initio} calculations of critical physical processes in galaxy evolution, such as star formation, supernova explosions, or accretion onto black holes.  Instead, they include physically motivated {\it sub-grid} modeling schemes that attempt to capture the key features of the underlying mechanisms. As these sub-grid models are not tuned to reproduce CGM and ICM observations as a function of mass and redshift, they make meaningful predictions, which differ substantially between models~\citep{Hillinprep}.  New, high signal-to-noise (S/N) measurements of the CGM and ICM will directly test these sub-grid feedback models and/or inform phenomenological semi-analytic prescriptions \citep[e.g.,][]{benson_galacticus:_2012}. Amongst the broad range of astrophysical implications, these measurements will calibrate the importance of stellar versus AGN feedback at different masses and redshifts, the roles of ejective versus preventative feedback, and the importance of radiatively inefficient (jet mode) versus radiatively efficient (i.e., radiation and winds) feedback.

\vspace{-6mm}

\section{New Tools from Old Photons}
\vspace{-2mm}

As CMB photons propagate through the Universe, they are scattered by free electrons in the ICM and CGM, resulting in secondary temperature fluctuations that dominate the anisotropy on arcminute scales \citep{SZ1969,SZ1970}. These secondary anisotropies, known as the Sunyaev--Zel'dovich effects~\citep{SZ1969,SZ1970,SZ1972}, contain a wealth of information about the abundance and thermodynamic history of baryons across cosmic time, as well as the growth of structure.  The thermal SZ (tSZ) effect is the inverse-Compton scattering of CMB photons off hot electrons, which results in a unique CMB spectral distortion. The tSZ signal is proportional to the integrated electron pressure along the line-of-sight. The kinematic SZ (kSZ) effect is the Doppler shift of CMB photons that Thomson-scatter off free electrons with a non-zero peculiar velocity with respect to the CMB rest frame.  To lowest order, the kSZ effect preserves the CMB blackbody spectrum, producing a CMB temperature shift proportional to the peculiar momentum of the electrons along the line-of-sight.  {\bf Both the tSZ and kSZ signals contain information about the thermodynamic properties of free electrons in the CGM and ICM.}  In particular, the kSZ signal directly traces the distribution of baryons in the CGM, which have generally eluded detection (``missing baryons'')~\citep[e.g.,][]{Bregman2007}.

\begin{figure}[t]
\includegraphics[width=0.5\textwidth]{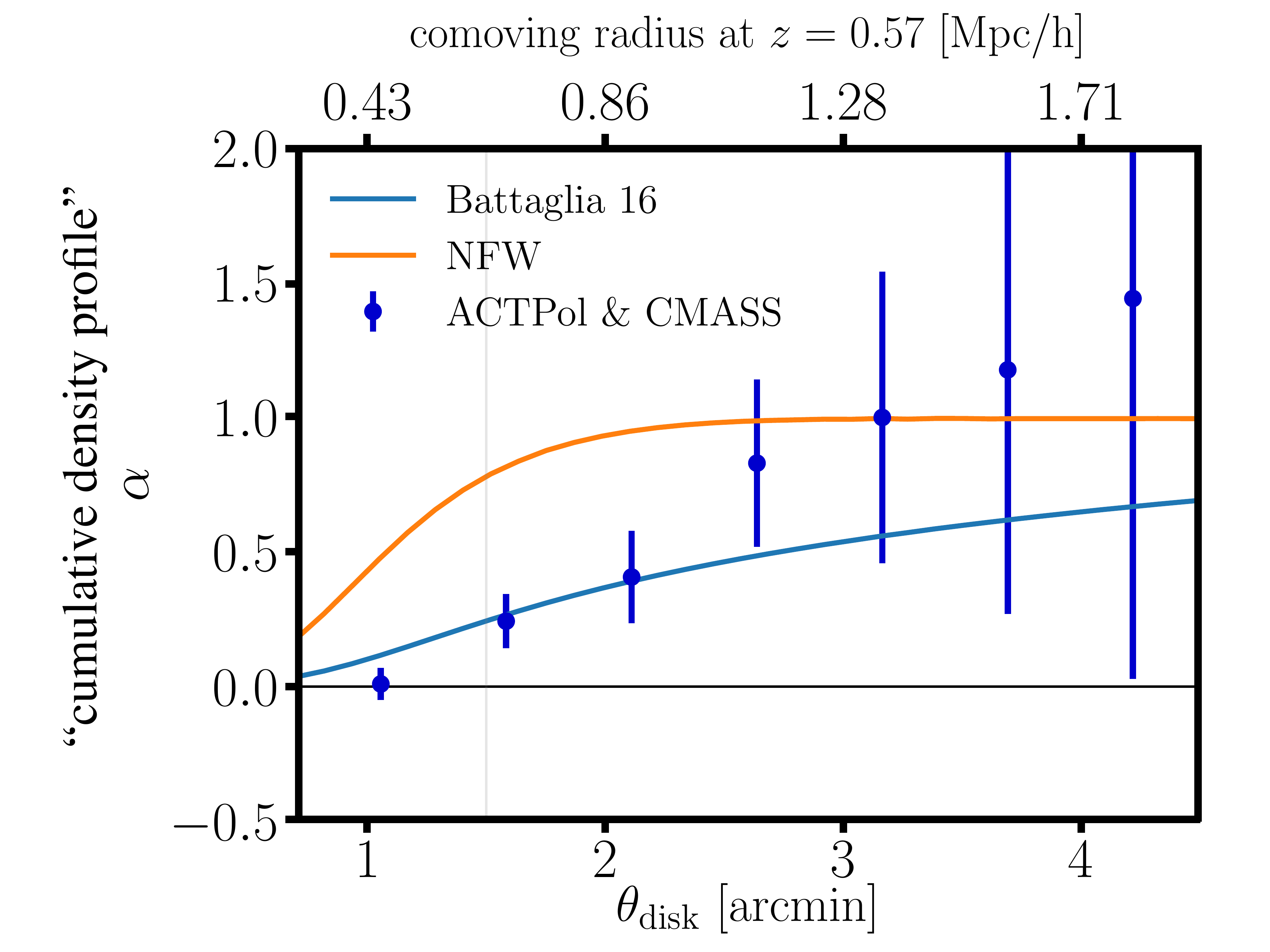}
\includegraphics[width=0.5\textwidth]{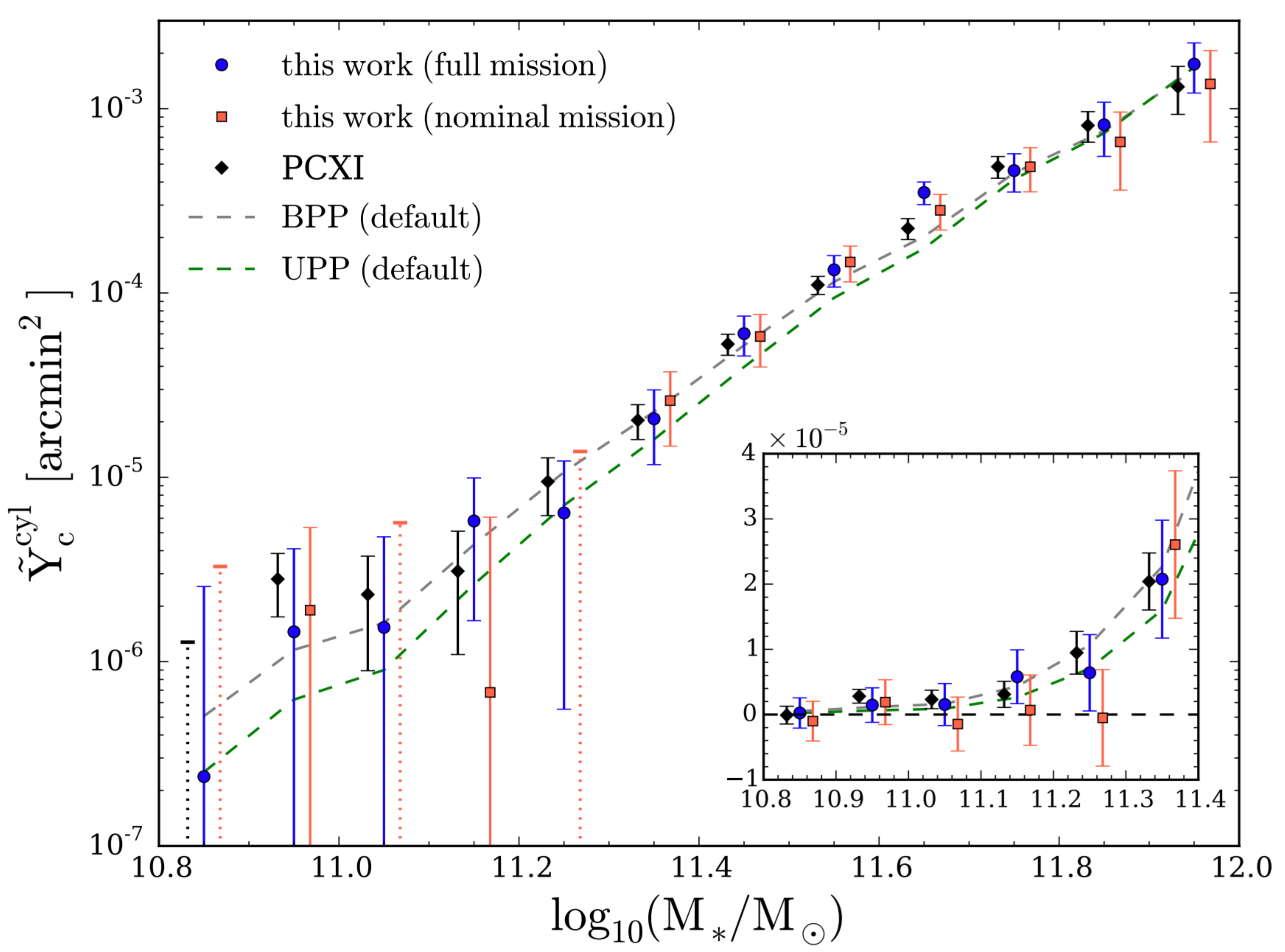}
\caption{\footnotesize {\it Left}: Measured kSZ signal from BOSS CMASS galaxies in ACTPol CMB maps \citep{Schaan2016}. The data points show a proxy for the cumulative gas mass fraction as a function of cluster-centric radius (in arcmin, or comoving Mpc$/h$ at $z=0.57$).  A model where gas follows dark matter (NFW, orange curve) is less favored than predictions from hydrodynamic simulations that include feedback \citep[blue curve,][]{Batt2016}. {\it Right}: Stacked tSZ signal of ``locally brightest galaxies'' (extracted from SDSS) as a function of their stellar mass, measured from Planck data~\citep{Greco2015} \citep[see also][]{Planckstack}.  The results are consistent with self-similar expectations (dashed curves), but this is a consequence of the coarse resolution of Planck~\cite{LeBrun2015,Greco2015,Vinu2018,Hill2018}; upcoming high-resolution data will probe the inner regions of these halos, where feedback is expected to produce deviations from self-similarity.
\vspace{-10mm}}
\label{fig:kSZ}
\end{figure}

Through cross-correlations of the CMB with galaxy samples, the tSZ signal has been detected down to nearly Milky Way-sized galaxies, with stellar masses of $M_* \approx 10^{11} \, \mathrm{M}_{\odot}$~\citep{Planckstack,Greco2015} (see Fig.~\ref{fig:kSZ}, right panel).  Such measurements have probed the average total thermal energy (integrated Compton-$y$) of the galaxy sample within a particular radial aperture (e.g., $5 r_{500}$ for the {\it Planck}-based measurements).  Similar to X-ray luminosity scaling relations, the integrated Compton-$y$ is sensitive to the thermal history of the galaxies and has been used to place constraints on feedback processes that inject additional energy into these halos \citep{Spacek2016,Spacek2017}.

The kSZ signal was first detected in 2012 \citep{Hand2012}.  Multiple detections have since followed, applying various estimators to CMB and galaxy data sets \citep[e.g.,][]{PlanckkSZ,Schaan2016,Hill2016,SPTkSZ2016,FDB2016} (e.g., Fig.~\ref{fig:kSZ}, left panel), as well as detections from individual galaxy clusters \citep{Sayers2013,Adam2017}. While the highest significance to date is $\approx 4\sigma$, forecasts for experiments like the Advanced Atacama Cosmology Telescope~\citep[AdvACT,][]{advact}, South Pole Telescope-3G~\citep[SPT-3G,][]{SPT3G}, Simons Observatory~\citep[SO,][]{SOforecasts}, and CMB Stage-4~\citep[CMB-S4,][]{CMBS4} promise S/N improvements of $\gtrsim$ 10--100~\citep{F16,BFSS2017}.

The next step is to measure the radial profiles of the tSZ and kSZ signals at high S/N as the shapes of the thermodynamic profiles (gas density, pressure, and temperature) are sensitive to the underlying physical processes that govern the CGM and ICM \citep[e.g.,][]{OBB2005,Nagai2007,BBPSS2010,LeBrun2014}. It has already been demonstrated that combining measurements of thermodynamic profiles from kSZ and tSZ observations places tight constraints on baryonic processes like feedback and non-thermal pressure support in the CGM and ICM \citep[e.g.,][]{BFSS2017}.  Future stacked SZ observations have the potential to probe gas out to the virial radii of halos down to Milky Way masses ($M_{200} \gtrsim 10^{12} \, M_{\odot}$) throughout the epochs of galaxy and cluster formation (see Fig.~\ref{fig:ill}, left panel); Stage-3 and Stage-4 CMB experiments are required to realize this goal. It should be possible to measure the ``splashback'' \citep[e.g.,][]{Lau2015} feature in SZ data and constrain internal CGM kinematics (e.g., rotation or turbulence) using high-resolution kSZ data.

On larger scales ($\gtrsim$ few Mpc), the tSZ and kSZ signals receive additional contributions from correlated galaxies or clusters, often referred to as the ``two-halo'' contribution. These contributions contain information on the average thermodynamic properties of the intergalactic medium (IGM). The distribution and thermodynamic properties of the IGM at large halo-centric radii \citep[e.g., ``missing baryons''][]{Bregman2007} remain open questions.  The tSZ two-halo term has been measured and shown to be important in halos of mass $\lesssim 10^{13}M_\odot$ \citep{Vinu2018,Hill2018}. Additionally, the tSZ signal of filaments has recently been measured, opening up another opportunity to probe the IGM between galaxy pairs \citep{Tanimura2019,deGraaff2017}. For the kSZ signal, the two-halo contributions are expected to come from halos within 50 Mpc, which is roughly the correlation length for the linear velocity field. The ``projected-field’’ kSZ estimator \citep{Dore2004,Hill2016,F16}, which allows for larger samples of photometrically selected galaxies to be used, probes the two-halo regime and has placed limits on the abundance of baryons in the IGM at the $4\sigma$ level.  The next decade will yield order-of-magnitude gains in S/N using this approach, by combining ground-based CMB experiments with large-area optical surveys \citep{F16}.

The tSZ and kSZ surface brightness is redshift-independent. Thus, the same SZ observations that are used to measure the thermodynamic properties of galaxies at low redshift can be and have been made using samples of high-redshift objects (e.g., quasars), which are very difficult to probe otherwise. For quasar samples, the tSZ signal has been measured at $\approx 4\sigma$ \citep{Ruan2015,Verdier2016,Crich2016,Soergel2016}; it has been argued that these observations provide evidence for non-gravitational heating (i.e., feedback) in the gas in these systems. As larger high-redshift samples become available from LSST and DESI, and as CMB data improve in sensitivity \citep[from the ground and from space, e.g., PICO][]{PICO}, we expect a large increase in S/N and an order-of magnitude-improvement over our current understanding of feedback efficiency and the thermodynamic properties of these halos over a wide range of redshift \citep[e.g.,][]{BFSS2017}.  {\bf These gains are driven by the dramatic improvement in sensitivity of ongoing and upcoming CMB experiments, including multi-frequency coverage; the pioneering SZ data of {\em Planck} represent only the beginning of an exciting scientific journey.}

\vspace{-6mm}
\section{Synergies Across the Electromagnetic Spectrum}
\vspace{-2mm}

\noindent {\underline{\bf X-rays}}: For galaxy clusters at low to moderate redshift, high-resolution X-ray observations from the {\it Chandra} and {\it XMM-Newton} satellites have revealed the complex interplay between thermodynamic and chemical properties of the ICM and the AGN of the central cluster galaxy out to the virial radius \citep[e.g.,][]{Mac2007}. Further improvements are imminent with the launch of {\it eROSITA}, followed by the future X-ray satellites {\it Athena} and {\it Lynx} in the 2030s, which will probe the thermodynamic state and chemical composition of the CGM with high-resolution to a large fraction of the virial radius, for systems out to $z \lesssim 1.5$. Like the SZ observations described above, these measurements will probe the effects of assembly history and feedback that shape galaxy formation. Such observations will be independent and complementary to those from the millimeter band with respect to radial range, halo mass, redshift coverage, and other properties (e.g., mass-weighted vs. emission-weighted temperature).

\noindent {\underline{\bf Absorption Line Studies}}: Targeted spectroscopic campaigns using background galaxies and quasars that intersect foreground galaxy halos are another technique used to assess the CGM through absorption-line analysis along these lines-of-sight~\citep[e.g.,][]{Tumlinson-COS-Halos,Chen2010,Rudie2019}. These measurements have shown that a substantial amount of cool, photoionized gas exists in the CGM for a variety of galaxies with different masses and star-formation histories. The abundance of this cool gas, not observable in X-rays or tSZ, can be inferred via detailed theoretical modeling, including assumptions that the CGM is in local thermal equilibrium with the extragalactic ultraviolet background. Future sample sizes for these analyses will grow in the next decade, helping to overcome a limiting factor for inferring average CGM properties.  These studies are highly complementary in halo mass and redshift range to the tSZ and kSZ probes, and measure a physically distinct gas phase.  Also, the tSZ and kSZ signals directly measure the electron pressure and density, rather than tracer ion populations, and thus are not affected by uncertainties in the metallicity and ionization state.

\noindent {\underline{\bf Fast Radio Bursts}}: Fast radio bursts (FRBs) probe the line-of-sight density of free electrons via frequency-dependent delays induced in the FRB signal, known as the ``dispersion measure'' (DM).  It has been shown theoretically that the cross-correlation of FRB DMs with galaxy surveys can constrain the baryon distribution in halos~\citep[e.g.,][]{McQuinn2014,Kiyo2015,Fujita2017,Ravi2018,ML2018,MatFRB2019}. Similar to kSZ observations, the electron profiles inferred from FRB DMs are unbiased. The reality of using FRB DMs to make such measurements is not currently viable as this requires three to four orders of magnitude more FRB detections than the $\lesssim 3$ currently available (with localization and redshift information). Dedicated radio-wave survey experiments coming online with factors of a few increase in size, such as CHIME~\citep{CHIME2018}, HIRAX~\citep{HIRAX2016}, or DSA-110, could feasibly push the number of FRB detections to $\approx 10^6$ events in the next decade.

\vspace{-6mm}
\section{Outlook for the Next Decade}
\vspace{-2mm}
\begin{figure}[t]
\includegraphics[width=0.58\textwidth]{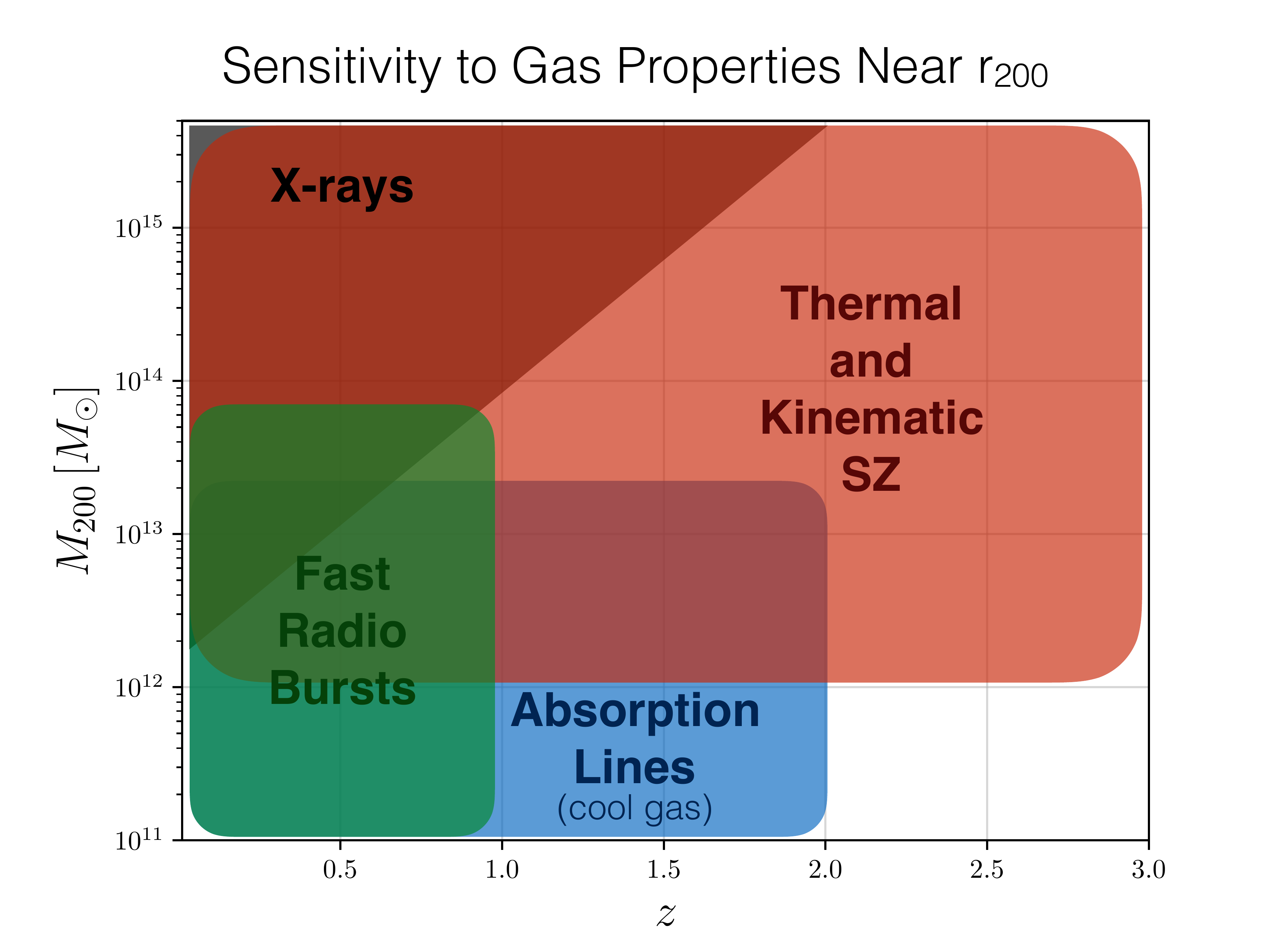}\includegraphics[width=0.42\textwidth]{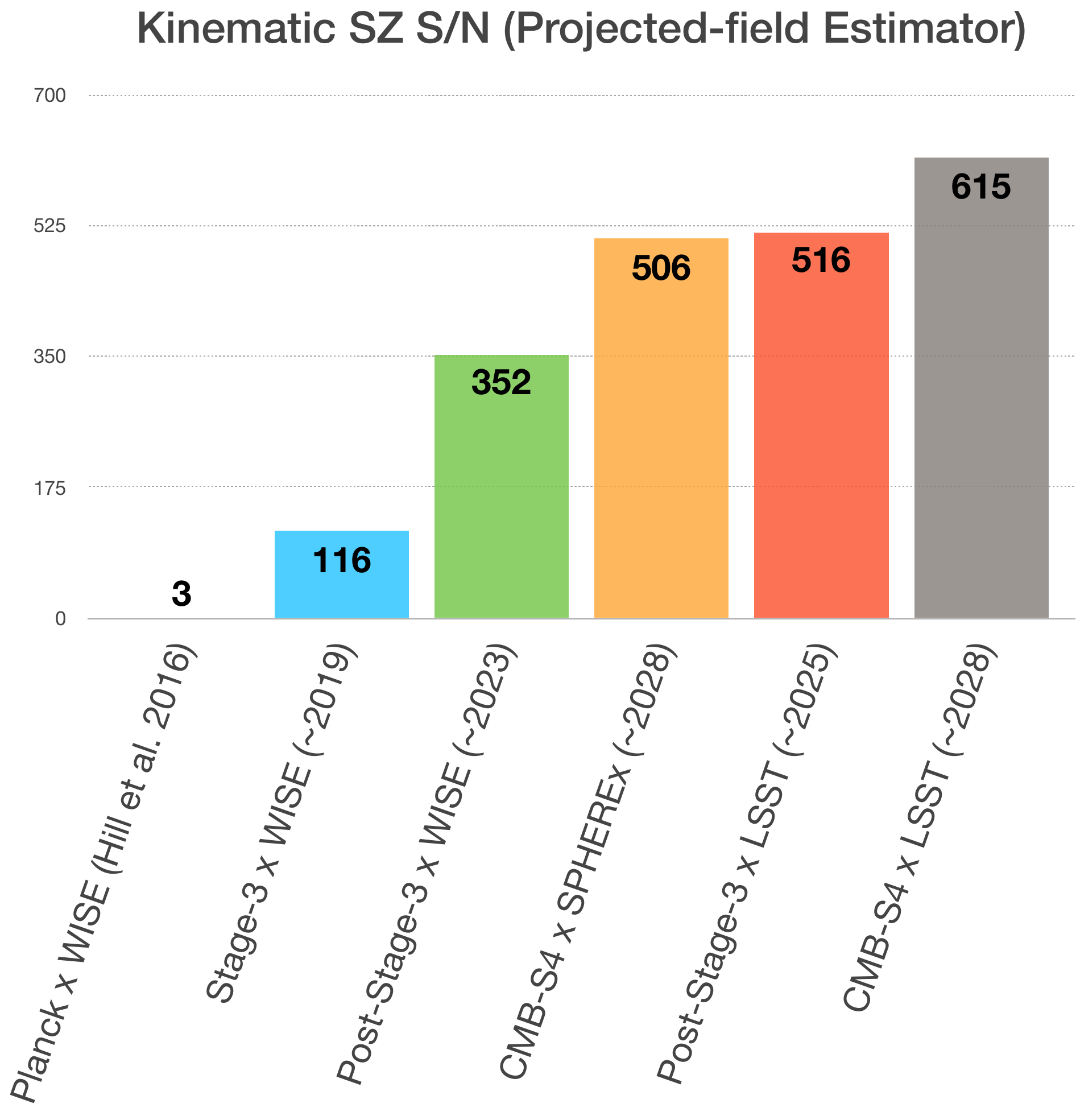}
   \caption{\footnotesize {\it Left}: Illustration of (approximate) mass and redshift ranges that tSZ/kSZ (red), X-ray (gray), FRB (green), and absorption line (blue) probes of the CGM and ICM will cover over the next decade.  We focus on properties near $r_{200}$ here, but most probes will also yield radial information.  The sensitivities assume the existence of large halo catalogs, as will be extracted from LSST data (and via tSZ out to high redshifts).  {\it Right}: Forecast S/N for the next decade of kSZ measurements using the ``projected-field'' estimator~\citep{Hill2016,F16}.
   \vspace{-10mm}
   }
\label{fig:ill}
\end{figure}

We are poised to make great strides in understanding the physical processes that govern galaxy formation through multi-wavelength observations of the CGM and ICM across broad spatial, mass, and redshift ranges (see Fig.~\ref{fig:ill}).  Amongst these probes, high-fidelity, high-resolution tSZ and kSZ observations are powerful, newly emergent tools.

These new tSZ and kSZ measurements encode information on the effects of assembly history and feedback processes that shape galaxy and cluster formation. In particular, they will provide critical information on the nature of feedback mechanisms in the CGM and ICM, which is required to narrow down the space of structure formation models.  They will directly test the sub-grid feedback models used in state-of-the-art cosmological hydrodynamic simulations and inform phenomenological semi-analytic prescriptions for galaxy formation, which both reproduce the optical properties of galaxies as currently measured.

Moreover, these measurements will play a crucial role in pinning down a major source of theoretical uncertainty for upcoming cosmological weak lensing surveys, namely, baryonic effects on the small-scale matter power spectrum~\citep[e.g.,][]{Schneider2015}.  Upcoming cosmological surveys, including LSST, will use measurements of weak lensing \citep[e.g.,][]{Shearreview2015} into the mildly non-linear regime, where there is ambiguity between the impact of cosmological parameters and physical processes that govern baryons in galaxies and clusters. Understanding these systematic effects from galaxy formation, and disentangling them from cosmological signals, is a major challenge, in which tSZ and kSZ observations are poised to play a key role.

These new tools are enabled by arcminute-resolution, Stage-3 CMB experiments on the sky now \citep[AdvACT and SPT-3G][]{advact,SPT3G}, as well as new telescopes coming online at the start of the decade \citep[SO and CCAT-prime][]{simonsproj,SOforecasts,CCATP}, CMB-S4 \citep{CMBS4} starting in the middle of the decade, and a possible Probe-class space mission starting late in the decade, e.g., PICO \citep{PICO}.  In addition, there are prospects for tSZ and kSZ measurements with even higher angular resolution using  $\gtrsim 30$m-scale telescopes like AtLAST\citep{atlast,Hargrave2018}, LST\citep{Kawabe2016}, or CSST\cite{Padin2014,Golwala2018}, and in smaller fields-of-view with NIKA2\cite{NIKA2} and ALMA\citep{Molnar2009,Kitayama2016}. {\bf Vast cosmological and astrophysical information will be obtained from these transformative measurements.  We recommend broad support for the range of experimental, observational, and theoretical work in this area in the coming decade.}

\clearpage
\bibliographystyle{unsrturltrunc6}
\bibliography{references.bib}

\begin{thebibliography}{10}

\bibitem{Pillepich2018}
Annalisa {Pillepich}, Dylan {Nelson}, Lars {Hernquist}, Volker {Springel},
  R{\"u}diger {Pakmor}, Paul {Torrey}, et~al.
\newblock {First results from the IllustrisTNG simulations: the stellar mass
  content of groups and clusters of galaxies}.
\newblock {\em \mnras}, 475:648--675, Mar 2018.
\newblock \href {http://arxiv.org/abs/1707.03406} {\path{arXiv:1707.03406}},
  \href {http://dx.doi.org/10.1093/mnras/stx3112}
  {\path{doi:10.1093/mnras/stx3112}}.

\bibitem{vandeVoort2016}
Freeke {van de Voort}, Eliot {Quataert}, Philip~F. {Hopkins}, Claude-Andr{\'e}
  {Faucher-Gigu{\`e}re}, Robert {Feldmann}, Du{\v{s}}an {Kere{\v{s}}}, et~al.
\newblock {The impact of stellar feedback on hot gas in galaxy haloes: the
  Sunyaev-Zel'dovich effect and soft X-ray emission}.
\newblock {\em \mnras}, 463:4533--4544, Dec 2016.
\newblock \href {http://arxiv.org/abs/1604.01397} {\path{arXiv:1604.01397}},
  \href {http://dx.doi.org/10.1093/mnras/stw2322}
  {\path{doi:10.1093/mnras/stw2322}}.

\bibitem{Hillinprep}
J.~C. {Hill}, N.~B. {Battaglia}, and et~al.
\newblock {Distinguishing Between Feedback Models with Thermal and Kinematic SZ
  Observations}.
\newblock {\em in prep.}, 2019.

\bibitem{RS2015}
R.~S. {Somerville} and R.~{Dav{\'e}}.
\newblock {Physical Models of Galaxy Formation in a Cosmological Framework}.
\newblock {\em \araa}, 53:51--113, August 2015.
\newblock \href {http://arxiv.org/abs/1412.2712} {\path{arXiv:1412.2712}},
  \href {http://dx.doi.org/10.1146/annurev-astro-082812-140951}
  {\path{doi:10.1146/annurev-astro-082812-140951}}.

\bibitem{fire}
Philip~F. {Hopkins}, Du{\v{s}}an {Kere{\v{s}}}, Jos{\'e} {O{\~n}orbe},
  Claude-Andr{\'e} {Faucher-Gigu{\`e}re}, Eliot {Quataert}, Norman {Murray},
  et~al.
\newblock {Galaxies on FIRE (Feedback In Realistic Environments): stellar
  feedback explains cosmologically inefficient star formation}.
\newblock {\em \mnras}, 445:581--603, Nov 2014.
\newblock \href {http://arxiv.org/abs/1311.2073} {\path{arXiv:1311.2073}},
  \href {http://dx.doi.org/10.1093/mnras/stu1738}
  {\path{doi:10.1093/mnras/stu1738}}.

\bibitem{Eagles}
J.~{Schaye}, R.~A. {Crain}, R.~G. {Bower}, M.~{Furlong}, M.~{Schaller},
  T.~{Theuns}, et~al.
\newblock {The EAGLE project: simulating the evolution and assembly of galaxies
  and their environments}.
\newblock {\em \mnras}, 446:521--554, January 2015.
\newblock \href {http://arxiv.org/abs/1407.7040} {\path{arXiv:1407.7040}},
  \href {http://dx.doi.org/10.1093/mnras/stu2058}
  {\path{doi:10.1093/mnras/stu2058}}.

\bibitem{HorizonAGN}
S.~{Kaviraj}, C.~{Laigle}, T.~{Kimm}, J.~E.~G. {Devriendt}, Y.~{Dubois},
  C.~{Pichon}, et~al.
\newblock {The Horizon-AGN simulation: evolution of galaxy properties over
  cosmic time}.
\newblock {\em \mnras}, 467:4739--4752, June 2017.
\newblock \href {http://arxiv.org/abs/1605.09379} {\path{arXiv:1605.09379}},
  \href {http://dx.doi.org/10.1093/mnras/stx126}
  {\path{doi:10.1093/mnras/stx126}}.

\bibitem{tng2018}
V.~{Springel}, R.~{Pakmor}, A.~{Pillepich}, R.~{Weinberger}, D.~{Nelson},
  L.~{Hernquist}, et~al.
\newblock {First results from the IllustrisTNG simulations: matter and galaxy
  clustering}.
\newblock {\em \mnras}, 475:676--698, March 2018.
\newblock \href {http://arxiv.org/abs/1707.03397} {\path{arXiv:1707.03397}},
  \href {http://dx.doi.org/10.1093/mnras/stx3304}
  {\path{doi:10.1093/mnras/stx3304}}.

\bibitem{benson_galacticus:_2012}
Andrew~J. Benson.
\newblock {GALACTICUS}: {A} semi-analytic model of galaxy formation.
\newblock {\em New Astronomy}, 17:175--197, February 2012.
\newblock URL: \url{http://adsabs.harvard.edu/abs/2012NewA...17..175B}.

\bibitem{SZ1969}
R.~A. {Sunyaev} and Y.~B. {Zel'dovich}.
\newblock {Distortions of the Background Radiation Spectrum}.
\newblock {\em \nat}, 223:721--722, August 1969.
\newblock \href {http://dx.doi.org/10.1038/223721a0}
  {\path{doi:10.1038/223721a0}}.

\bibitem{SZ1970}
R.~A. {Sunyaev} and Y.~B. {Zel'dovich}.
\newblock {Small-Scale Fluctuations of Relic Radiation}.
\newblock 7:3, 1970.
\newblock URL:
  \url{http://adsabs.harvard.edu/cgi-bin/nph-bib_query?bibcode=1970Ap%26SS...7....3S&db_key=AST}.

\bibitem{SZ1972}
R.~A. {Sunyaev} and Y.~B. {Zel'dovich}.
\newblock {The Observations of Relic Radiation as a Test of the Nature of X-Ray
  Radiation from the Clusters of Galaxies}.
\newblock {\em Comments on Astrophysics and Space Physics}, 4:173, November
  1972.

\bibitem{Bregman2007}
J.~N. {Bregman}.
\newblock {The Search for the Missing Baryons at Low Redshift}.
\newblock {\em \araa}, 45:221--259, September 2007.
\newblock \href {http://arxiv.org/abs/0706.1787} {\path{arXiv:0706.1787}},
  \href {http://dx.doi.org/10.1146/annurev.astro.45.051806.110619}
  {\path{doi:10.1146/annurev.astro.45.051806.110619}}.

\bibitem{Planckstack}
{Planck Collaboration}, P.~A.~R. {Ade}, N.~{Aghanim}, M.~{Arnaud},
  M.~{Ashdown}, F.~{Atrio-Barandela}, et~al.
\newblock {Planck intermediate results. XI. The gas content of dark matter
  halos: the Sunyaev-Zeldovich-stellar mass relation for locally brightest
  galaxies}.
\newblock {\em \aap}, 557:A52, September 2013.
\newblock \href {http://arxiv.org/abs/1212.4131} {\path{arXiv:1212.4131}},
  \href {http://dx.doi.org/10.1051/0004-6361/201220941}
  {\path{doi:10.1051/0004-6361/201220941}}.

\bibitem{Greco2015}
J.~P. {Greco}, J.~C. {Hill}, D.~N. {Spergel}, and N.~{Battaglia}.
\newblock {The Stacked Thermal Sunyaev-Zel'dovich Signal of Locally Brightest
  Galaxies in Planck Full Mis\ sion Data: Evidence for Galaxy Feedback?}
\newblock {\em \apj}, 808:151, August 2015.
\newblock \href {http://arxiv.org/abs/1409.6747} {\path{arXiv:1409.6747}},
  \href {http://dx.doi.org/10.1088/0004-637X/808/2/151}
  {\path{doi:10.1088/0004-637X/808/2/151}}.

\bibitem{Spacek2016}
A.~{Spacek}, E.~{Scannapieco}, S.~{Cohen}, B.~{Joshi}, and P.~{Mauskopf}.
\newblock {Constraining AGN Feedback in Massive Ellipticals with South Pole
  Telescope Measurements of the Thermal Sunyaev-Zel'dovich Effect}.
\newblock {\em \apj}, 819:128, March 2016.
\newblock \href {http://arxiv.org/abs/1601.01330} {\path{arXiv:1601.01330}},
  \href {http://dx.doi.org/10.3847/0004-637X/819/2/128}
  {\path{doi:10.3847/0004-637X/819/2/128}}.

\bibitem{Spacek2017}
A.~{Spacek}, E.~{Scannapieco}, S.~{Cohen}, B.~{Joshi}, and P.~{Mauskopf}.
\newblock {Searching for Fossil Evidence of AGN Feedback in WISE-selected
  Stripe-82 Galaxies by Measuring the Thermal Sunyaev{\ndash}Zel{\rsquo}dovich
  Effect with the Atacama Cosmology Telescope}.
\newblock {\em \apj}, 834:102, January 2017.
\newblock \href {http://arxiv.org/abs/1610.02068} {\path{arXiv:1610.02068}},
  \href {http://dx.doi.org/10.3847/1538-4357/834/2/102}
  {\path{doi:10.3847/1538-4357/834/2/102}}.

\bibitem{Hand2012}
N.~{Hand}, G.~E. {Addison}, E.~{Aubourg}, N.~{Battaglia}, E.~S. {Battistelli},
  D.~{Bizyaev}, et~al.
\newblock {Evidence of Galaxy Cluster Motions with the Kinematic
  Sunyaev-Zel'dovich Effect}.
\newblock {\em Physical Review Letters}, 109(4):041101, July 2012.
\newblock \href {http://arxiv.org/abs/1203.4219} {\path{arXiv:1203.4219}},
  \href {http://dx.doi.org/10.1103/PhysRevLett.109.041101}
  {\path{doi:10.1103/PhysRevLett.109.041101}}.

\bibitem{PlanckkSZ}
{Planck Collaboration}, P.~A.~R. {Ade}, N.~{Aghanim}, M.~{Arnaud},
  M.~{Ashdown}, E.~{Aubourg}, et~al.
\newblock {Planck intermediate results. XXXVII. Evidence of unbound gas from
  the kinetic Sunyaev-Zeldovich effect}.
\newblock {\em \aap}, 586:A140, February 2016.
\newblock \href {http://arxiv.org/abs/1504.03339} {\path{arXiv:1504.03339}},
  \href {http://dx.doi.org/10.1051/0004-6361/201526328}
  {\path{doi:10.1051/0004-6361/201526328}}.

\bibitem{Schaan2016}
E.~{Schaan}, S.~{Ferraro}, M.~{Vargas-Maga{\~n}a}, K.~M. {Smith}, S.~{Ho},
  S.~{Aiola}, et~al.
\newblock {Evidence for the kinematic Sunyaev-Zel'dovich effect with the
  Atacama Cosmology Telescope and velocity reconstruction from the Baryon
  Oscillation Spectroscopic Survey}.
\newblock {\em \prd}, 93(8):082002, April 2016.
\newblock \href {http://dx.doi.org/10.1103/PhysRevD.93.082002}
  {\path{doi:10.1103/PhysRevD.93.082002}}.

\bibitem{Hill2016}
J.~Colin {Hill}, Simone {Ferraro}, Nick {Battaglia}, Jia {Liu}, and David~N.
  {Spergel}.
\newblock {Kinematic Sunyaev-Zel'dovich Effect with Projected Fields: A Novel
  Probe of the Baryon Distribution with Planck, WMAP, and WISE Data}.
\newblock {\em \prl}, 117:051301, Jul 2016.
\newblock \href {http://arxiv.org/abs/1603.01608} {\path{arXiv:1603.01608}},
  \href {http://dx.doi.org/10.1103/PhysRevLett.117.051301}
  {\path{doi:10.1103/PhysRevLett.117.051301}}.

\bibitem{SPTkSZ2016}
B.~{Soergel}, S.~{Flender}, K.~T. {Story}, L.~{Bleem}, T.~{Giannantonio},
  G.~{Efstathiou}, et~al.
\newblock {Detection of the kinematic Sunyaev-Zel'dovich effect with DES Year 1
  and SPT}.
\newblock {\em \mnras}, 461:3172--3193, September 2016.
\newblock \href {http://arxiv.org/abs/1603.03904} {\path{arXiv:1603.03904}},
  \href {http://dx.doi.org/10.1093/mnras/stw1455}
  {\path{doi:10.1093/mnras/stw1455}}.

\bibitem{FDB2016}
F.~{De Bernardis}, S.~{Aiola}, E.~M. {Vavagiakis}, N.~{Battaglia}, M.~D.
  {Niemack}, J.~{Beall}, et~al.
\newblock {Detection of the pairwise kinematic Sunyaev-Zel'dovich effect with
  BOSS DR11 and the Atacama Cosmology Telescope}.
\newblock {\em \jcap}, 3:008, March 2017.
\newblock \href {http://arxiv.org/abs/1607.02139} {\path{arXiv:1607.02139}},
  \href {http://dx.doi.org/10.1088/1475-7516/2017/03/008}
  {\path{doi:10.1088/1475-7516/2017/03/008}}.

\bibitem{Sayers2013}
J.~{Sayers}, T.~{Mroczkowski}, M.~{Zemcov}, P.~M. {Korngut}, J.~{Bock},
  E.~{Bulbul}, et~al.
\newblock {A Measurement of the Kinetic Sunyaev-Zel'dovich Signal Toward MACS
  J0717.5+3745}.
\newblock {\em \apj}, 778:52, November 2013.
\newblock \href {http://arxiv.org/abs/1312.3680} {\path{arXiv:1312.3680}},
  \href {http://dx.doi.org/10.1088/0004-637X/778/1/52}
  {\path{doi:10.1088/0004-637X/778/1/52}}.

\bibitem{Adam2017}
R.~{Adam}, I.~{Bartalucci}, G.~W. {Pratt}, P.~{Ade}, P.~{Andr{\'e}},
  M.~{Arnaud}, et~al.
\newblock {Mapping the kinetic Sunyaev-Zel'dovich effect toward MACS
  J0717.5+3745 with NIKA}.
\newblock {\em \aap}, 598:A115, February 2017.
\newblock \href {http://arxiv.org/abs/1606.07721} {\path{arXiv:1606.07721}},
  \href {http://dx.doi.org/10.1051/0004-6361/201629182}
  {\path{doi:10.1051/0004-6361/201629182}}.

\bibitem{advact}
S.~W. {Henderson}, R.~{Allison}, J.~{Austermann}, T.~{Baildon}, N.~{Battaglia},
  J.~A. {Beall}, et~al.
\newblock {Advanced ACTPol Cryogenic Detector Arrays and Readout}.
\newblock {\em Journal of Low Temperature Physics}, 184:772--779, August 2016.
\newblock \href {http://arxiv.org/abs/1510.02809} {\path{arXiv:1510.02809}},
  \href {http://dx.doi.org/10.1007/s10909-016-1575-z}
  {\path{doi:10.1007/s10909-016-1575-z}}.

\bibitem{SPT3G}
B.~A. {Benson}, P.~A.~R. {Ade}, Z.~{Ahmed}, S.~W. {Allen}, K.~{Arnold}, J.~E.
  {Austermann}, et~al.
\newblock {SPT-3G: a next-generation cosmic microwave background polarization
  experiment on the South Pole telescope}.
\newblock In {\em Millimeter, Submillimeter, and Far-Infrared Detectors and
  Instrumentation for Astronomy VII}, volume 9153 of {\em \procspie}, page
  91531P, July 2014.
\newblock \href {http://arxiv.org/abs/1407.2973} {\path{arXiv:1407.2973}},
  \href {http://dx.doi.org/10.1117/12.2057305} {\path{doi:10.1117/12.2057305}}.

\bibitem{SOforecasts}
P.~{Ade}, J.~{Aguirre}, Z.~{Ahmed}, S.~{Aiola}, A.~{Ali}, D.~{Alonso}, et~al.
\newblock {The Simons Observatory: science goals and forecasts}.
\newblock {\em \jcap}, 2:056, February 2019.
\newblock \href {http://arxiv.org/abs/1808.07445} {\path{arXiv:1808.07445}},
  \href {http://dx.doi.org/10.1088/1475-7516/2019/02/056}
  {\path{doi:10.1088/1475-7516/2019/02/056}}.

\bibitem{CMBS4}
Kevork~N. {Abazajian}, Peter {Adshead}, Zeeshan {Ahmed}, Steven~W. {Allen},
  David {Alonso}, Kam~S. {Arnold}, et~al.
\newblock {CMB-S4 Science Book, First Edition}.
\newblock {\em arXiv e-prints}, page arXiv:1610.02743, October 2016.
\newblock \href {http://arxiv.org/abs/1610.02743} {\path{arXiv:1610.02743}}.

\bibitem{F16}
S.~{Ferraro}, J.~C. {Hill}, N.~{Battaglia}, J.~{Liu}, and D.~N. {Spergel}.
\newblock {Kinematic Sunyaev-Zel'dovich effect with projected fields. II.
  Prospects, challenges, and comparison with simulations}.
\newblock {\em \prd}, 94(12):123526, December 2016.
\newblock \href {http://arxiv.org/abs/1605.02722} {\path{arXiv:1605.02722}},
  \href {http://dx.doi.org/10.1103/PhysRevD.94.123526}
  {\path{doi:10.1103/PhysRevD.94.123526}}.

\bibitem{BFSS2017}
N.~{Battaglia}, S.~{Ferraro}, E.~{Schaan}, and D.~{Spergel}.
\newblock {Future constraints on halo thermodynamics from combined
  Sunyaev-Zel'dovich measurements}.
\newblock {\em ArXiv:1705.05881}, May 2017.
\newblock \href {http://arxiv.org/abs/1705.05881} {\path{arXiv:1705.05881}}.

\bibitem{Batt2016}
N.~{Battaglia}.
\newblock {The tau of galaxy clusters}.
\newblock {\em \jcap}, 8:058, August 2016.
\newblock \href {http://arxiv.org/abs/1607.02442} {\path{arXiv:1607.02442}},
  \href {http://dx.doi.org/10.1088/1475-7516/2016/08/058}
  {\path{doi:10.1088/1475-7516/2016/08/058}}.

\bibitem{LeBrun2015}
A.~M.~C. {Le Brun}, I.~G. {McCarthy}, and J.-B. {Melin}.
\newblock {Testing Sunyaev-Zel'dovich measurements of the hot gas content of
  dark matter haloes using synthetic skies}.
\newblock {\em \mnras}, 451:3868--3881, August 2015.
\newblock \href {http://arxiv.org/abs/1501.05666} {\path{arXiv:1501.05666}},
  \href {http://dx.doi.org/10.1093/mnras/stv1172}
  {\path{doi:10.1093/mnras/stv1172}}.

\bibitem{Vinu2018}
V.~{Vikram}, A.~{Lidz}, and B.~{Jain}.
\newblock {A Measurement of the Galaxy Group-Thermal Sunyaev-Zel'dovich Effect
  Cross-Correlation Function}.
\newblock {\em \mnras}, 467:2315--2330, May 2017.
\newblock \href {http://arxiv.org/abs/1608.04160} {\path{arXiv:1608.04160}},
  \href {http://dx.doi.org/10.1093/mnras/stw3311}
  {\path{doi:10.1093/mnras/stw3311}}.

\bibitem{Hill2018}
J.~C. {Hill}, E.~J. {Baxter}, A.~{Lidz}, J.~P. {Greco}, and B.~{Jain}.
\newblock {Two-halo term in stacked thermal Sunyaev-Zel'dovich measurements:
  Implications for self-similarity}.
\newblock {\em \prd}, 97(8):083501, April 2018.
\newblock \href {http://arxiv.org/abs/1706.03753} {\path{arXiv:1706.03753}},
  \href {http://dx.doi.org/10.1103/PhysRevD.97.083501}
  {\path{doi:10.1103/PhysRevD.97.083501}}.

\bibitem{OBB2005}
J.~P. {Ostriker}, P.~{Bode}, and A.~{Babul}.
\newblock {A Simple and Accurate Model for Intracluster Gas}.
\newblock {\em \apj}, 634:964--976, December 2005.
\newblock \href {http://arxiv.org/abs/arXiv:astro-ph/0504334}
  {\path{arXiv:arXiv:astro-ph/0504334}}, \href
  {http://dx.doi.org/10.1086/497122} {\path{doi:10.1086/497122}}.

\bibitem{Nagai2007}
D.~{Nagai}, A.~V. {Kravtsov}, and A.~{Vikhlinin}.
\newblock {Effects of Galaxy Formation on Thermodynamics of the Intracluster
  Medium}.
\newblock {\em \apj}, 668:1--14, October 2007.
\newblock \href {http://arxiv.org/abs/astro-ph/0703661}
  {\path{arXiv:astro-ph/0703661}}, \href {http://dx.doi.org/10.1086/521328}
  {\path{doi:10.1086/521328}}.

\bibitem{BBPSS2010}
N.~{Battaglia}, J.~R. {Bond}, C.~{Pfrommer}, J.~L. {Sievers}, and D.~{Sijacki}.
\newblock {Simulations of the Sunyaev-Zel'dovich Power Spectrum with Active
  Galactic Nucleus Feedback}.
\newblock {\em \apj}, 725:91--99, December 2010.
\newblock \href {http://arxiv.org/abs/1003.4256} {\path{arXiv:1003.4256}},
  \href {http://dx.doi.org/10.1088/0004-637X/725/1/91}
  {\path{doi:10.1088/0004-637X/725/1/91}}.

\bibitem{LeBrun2014}
A.~M.~C. {Le Brun}, I.~G. {McCarthy}, J.~{Schaye}, and T.~J. {Ponman}.
\newblock {Towards a realistic population of simulated galaxy groups and
  clusters}.
\newblock {\em \mnras}, 441:1270--1290, June 2014.
\newblock \href {http://arxiv.org/abs/1312.5462} {\path{arXiv:1312.5462}},
  \href {http://dx.doi.org/10.1093/mnras/stu608}
  {\path{doi:10.1093/mnras/stu608}}.

\bibitem{Lau2015}
E.~T. {Lau}, D.~{Nagai}, C.~{Avestruz}, K.~{Nelson}, and A.~{Vikhlinin}.
\newblock {Mass Accretion and its Effects on the Self-similarity of Gas
  Profiles in the Outskirts of Galaxy Clusters}.
\newblock {\em \apj}, 806:68, June 2015.
\newblock \href {http://arxiv.org/abs/1411.5361} {\path{arXiv:1411.5361}},
  \href {http://dx.doi.org/10.1088/0004-637X/806/1/68}
  {\path{doi:10.1088/0004-637X/806/1/68}}.

\bibitem{Tanimura2019}
Hideki {Tanimura}, Gary {Hinshaw}, Ian~G. {McCarthy}, Ludovic {Van Waerbeke},
  Nabila {Aghanim}, Yin-Zhe {Ma}, et~al.
\newblock {A search for warm/hot gas filaments between pairs of SDSS Luminous
  Red Galaxies}.
\newblock {\em \mnras}, 483:223--234, Feb 2019.
\newblock \href {http://arxiv.org/abs/1709.05024} {\path{arXiv:1709.05024}},
  \href {http://dx.doi.org/10.1093/mnras/sty3118}
  {\path{doi:10.1093/mnras/sty3118}}.

\bibitem{deGraaff2017}
A.~{de Graaff}, Y.-C. {Cai}, C.~{Heymans}, and J.~A. {Peacock}.
\newblock {Missing baryons in the cosmic web revealed by the Sunyaev-Zel'dovich
  effect}.
\newblock {\em arXiv e-prints}, September 2017.
\newblock \href {http://arxiv.org/abs/1709.10378} {\path{arXiv:1709.10378}}.

\bibitem{Dore2004}
O.~{Dor{\'e}}, J.~F. {Hennawi}, and D.~N. {Spergel}.
\newblock {Beyond the Damping Tail: Cross-Correlating the Kinetic
  Sunyaev-Zel'dovich Effect with Cosmic Shear}.
\newblock {\em \apj}, 606:46--57, May 2004.
\newblock \href {http://arxiv.org/abs/astro-ph/0309337}
  {\path{arXiv:astro-ph/0309337}}, \href {http://dx.doi.org/10.1086/382946}
  {\path{doi:10.1086/382946}}.

\bibitem{Ruan2015}
J.~J. {Ruan}, M.~{McQuinn}, and S.~F. {Anderson}.
\newblock {Detection of Quasar Feedback from the Thermal
  Sunyaev-Zel{\rsquo}dovich Effect in Planck}.
\newblock {\em \apj}, 802:135, April 2015.
\newblock \href {http://arxiv.org/abs/1502.01723} {\path{arXiv:1502.01723}},
  \href {http://dx.doi.org/10.1088/0004-637X/802/2/135}
  {\path{doi:10.1088/0004-637X/802/2/135}}.

\bibitem{Verdier2016}
L.~{Verdier}, J.-B. {Melin}, J.~G. {Bartlett}, C.~{Magneville},
  N.~{Palanque-Delabrouille}, and C.~{Y{\`e}che}.
\newblock {Quasar host environments: The view from Planck}.
\newblock {\em \aap}, 588:A61, April 2016.
\newblock \href {http://arxiv.org/abs/1509.07306} {\path{arXiv:1509.07306}},
  \href {http://dx.doi.org/10.1051/0004-6361/201527431}
  {\path{doi:10.1051/0004-6361/201527431}}.

\bibitem{Crich2016}
D.~{Crichton}, M.~B. {Gralla}, K.~{Hall}, T.~A. {Marriage}, N.~L. {Zakamska},
  N.~{Battaglia}, et~al.
\newblock {Evidence for the thermal Sunyaev-Zel'dovich effect associated with
  quasar feedback}.
\newblock {\em \mnras}, 458:1478--1492, May 2016.
\newblock \href {http://arxiv.org/abs/1510.05656} {\path{arXiv:1510.05656}},
  \href {http://dx.doi.org/10.1093/mnras/stw344}
  {\path{doi:10.1093/mnras/stw344}}.

\bibitem{Soergel2016}
B.~{Soergel}, T.~{Giannantonio}, G.~{Efstathiou}, E.~{Puchwein}, and
  D.~{Sijacki}.
\newblock {Constraints on AGN feedback from its Sunyaev-Zel'dovich imprint on
  the cosmic background radiation}.
\newblock {\em \mnras}, 468:577--596, June 2017.
\newblock \href {http://arxiv.org/abs/1612.06296} {\path{arXiv:1612.06296}},
  \href {http://dx.doi.org/10.1093/mnras/stx492}
  {\path{doi:10.1093/mnras/stx492}}.

\bibitem{PICO}
S.~{Hanany}, M.~{Alvarez}, E.~{Artis}, P.~{Ashton}, J.~{Aumont}, R.~{Aurlien},
  et~al.
\newblock {PICO: Probe of Inflation and Cosmic Origins}.
\newblock {\em arXiv e-prints}, February 2019.
\newblock \href {http://arxiv.org/abs/1902.10541} {\path{arXiv:1902.10541}}.

\bibitem{Mac2007}
B.~R. {McNamara} and P.~E.~J. {Nulsen}.
\newblock {Heating Hot Atmospheres with Active Galactic Nuclei}.
\newblock {\em \araa}, 45:117--175, September 2007.
\newblock \href {http://arxiv.org/abs/0709.2152} {\path{arXiv:0709.2152}},
  \href {http://dx.doi.org/10.1146/annurev.astro.45.051806.110625}
  {\path{doi:10.1146/annurev.astro.45.051806.110625}}.

\bibitem{Tumlinson-COS-Halos}
Jason {Tumlinson}, Christopher {Thom}, Jessica~K. {Werk}, J.~Xavier
  {Prochaska}, Todd~M. {Tripp}, Neal {Katz}, et~al.
\newblock {The COS-Halos Survey: Rationale, Design, and a Census of
  Circumgalactic Neutral Hydrogen}.
\newblock {\em \apj}, 777:59, Nov 2013.
\newblock \href {http://arxiv.org/abs/1309.6317} {\path{arXiv:1309.6317}},
  \href {http://dx.doi.org/10.1088/0004-637X/777/1/59}
  {\path{doi:10.1088/0004-637X/777/1/59}}.

\bibitem{Chen2010}
Hsiao-Wen {Chen}, Jennifer~E. {Helsby}, Jean-Ren{\'e} {Gauthier}, Stephen~A.
  {Shectman}, Ian~B. {Thompson}, and Jeremy~L. {Tinker}.
\newblock {An Empirical Characterization of Extended Cool Gas Around Galaxies
  Using Mg II Absorption Features}.
\newblock {\em \apj}, 714:1521--1541, May 2010.
\newblock \href {http://arxiv.org/abs/1004.0705} {\path{arXiv:1004.0705}},
  \href {http://dx.doi.org/10.1088/0004-637X/714/2/1521}
  {\path{doi:10.1088/0004-637X/714/2/1521}}.

\bibitem{Rudie2019}
G.~C. {Rudie}, C.~C. {Steidel}, M.~{Pettini}, R.~F. {Trainor}, A.~L. {Strom},
  C.~B. {Hummels}, et~al.
\newblock {The Column Density, Kinematics, and Thermal State of Metal-Bearing
  Gas within the Virial Radius of z\~{}2 Star-Forming Galaxies in the Keck
  Baryonic Structure Survey}.
\newblock {\em arXiv e-prints}, February 2019.
\newblock \href {http://arxiv.org/abs/1903.00004} {\path{arXiv:1903.00004}}.

\bibitem{McQuinn2014}
M.~{McQuinn}.
\newblock {Locating the ``Missing'' Baryons with Extragalactic Dispersion
  Measure Estimates}.
\newblock {\em The Astrophysical Journal Letters}, 780:L33, January 2014.
\newblock \href {http://arxiv.org/abs/1309.4451} {\path{arXiv:1309.4451}},
  \href {http://dx.doi.org/10.1088/2041-8205/780/2/L33}
  {\path{doi:10.1088/2041-8205/780/2/L33}}.

\bibitem{Kiyo2015}
Kiyoshi~Wesley {Masui} and Kris {Sigurdson}.
\newblock {Dispersion Distance and the Matter Distribution of the Universe in
  Dispersion Space}.
\newblock {\em \prl}, 115:121301, Sep 2015.
\newblock \href {http://arxiv.org/abs/1506.01704} {\path{arXiv:1506.01704}},
  \href {http://dx.doi.org/10.1103/PhysRevLett.115.121301}
  {\path{doi:10.1103/PhysRevLett.115.121301}}.

\bibitem{Fujita2017}
Y.~{Fujita}, T.~{Akahori}, K.~{Umetsu}, C.~L. {Sarazin}, and K.-W. {Wong}.
\newblock {Probing WHIM around Galaxy Clusters with Fast Radio Bursts and the
  Sunyaev-Zel{\rsquo}dovich effect}.
\newblock {\em \apj}, 834:13, January 2017.
\newblock \href {http://arxiv.org/abs/1609.03566} {\path{arXiv:1609.03566}},
  \href {http://dx.doi.org/10.3847/1538-4357/834/1/13}
  {\path{doi:10.3847/1538-4357/834/1/13}}.

\bibitem{Ravi2018}
Vikram {Ravi}.
\newblock {Measuring the Circumgalactic and Intergalactic Baryon Contents with
  Fast Radio Bursts}.
\newblock {\em \apj}, 872:88, Feb 2019.
\newblock \href {http://arxiv.org/abs/1804.07291} {\path{arXiv:1804.07291}},
  \href {http://dx.doi.org/10.3847/1538-4357/aafb30}
  {\path{doi:10.3847/1538-4357/aafb30}}.

\bibitem{ML2018}
Julian~B. {Mu{\~n}oz} and Abraham {Loeb}.
\newblock {Finding the missing baryons with fast radio bursts and
  Sunyaev-Zeldovich maps}.
\newblock {\em \prd}, 98:103518, Nov 2018.
\newblock \href {http://arxiv.org/abs/1809.04074} {\path{arXiv:1809.04074}},
  \href {http://dx.doi.org/10.1103/PhysRevD.98.103518}
  {\path{doi:10.1103/PhysRevD.98.103518}}.

\bibitem{MatFRB2019}
M.~S. {Madhavacheril}, N.~{Battaglia}, K.~M. {Smith}, and J.~L. {Sievers}.
\newblock {Cosmology with kSZ: breaking the optical depth degeneracy with Fast
  Radio Bursts}.
\newblock {\em arXiv e-prints}, January 2019.
\newblock \href {http://arxiv.org/abs/1901.02418} {\path{arXiv:1901.02418}}.

\bibitem{CHIME2018}
{CHIME/FRB Collaboration}, M.~{Amiri}, K.~{Bandura}, P.~{Berger},
  M.~{Bhardwaj}, M.~M. {Boyce}, et~al.
\newblock {The CHIME Fast Radio Burst Project: System Overview}.
\newblock {\em \apj}, 863:48, August 2018.
\newblock \href {http://arxiv.org/abs/1803.11235} {\path{arXiv:1803.11235}},
  \href {http://dx.doi.org/10.3847/1538-4357/aad188}
  {\path{doi:10.3847/1538-4357/aad188}}.

\bibitem{HIRAX2016}
L.~B. {Newburgh}, K.~{Bandura}, M.~A. {Bucher}, T.~C. {Chang}, H.~C. {Chiang},
  J.~F. {Cliche}, et~al.
\newblock {HIRAX: a probe of dark energy and radio transients}.
\newblock In {\em Ground-based and Airborne Telescopes VI}, volume 9906 of {\em
  Society of Photo-Optical Instrumentation Engineers (SPIE) Conference Series},
  page 99065X, Aug 2016.
\newblock \href {http://arxiv.org/abs/1607.02059} {\path{arXiv:1607.02059}},
  \href {http://dx.doi.org/10.1117/12.2234286} {\path{doi:10.1117/12.2234286}}.

\bibitem{Schneider2015}
Aurel {Schneider} and Romain {Teyssier}.
\newblock {A new method to quantify the effects of baryons on the matter power
  spectrum}.
\newblock {\em Journal of Cosmology and Astro-Particle Physics}, 2015:049, Dec
  2015.
\newblock \href {http://arxiv.org/abs/1510.06034} {\path{arXiv:1510.06034}},
  \href {http://dx.doi.org/10.1088/1475-7516/2015/12/049}
  {\path{doi:10.1088/1475-7516/2015/12/049}}.

\bibitem{Shearreview2015}
M.~{Kilbinger}.
\newblock {Cosmology with cosmic shear observations: a review}.
\newblock {\em Reports on Progress in Physics}, 78(8):086901, July 2015.
\newblock \href {http://arxiv.org/abs/1411.0115} {\path{arXiv:1411.0115}},
  \href {http://dx.doi.org/10.1088/0034-4885/78/8/086901}
  {\path{doi:10.1088/0034-4885/78/8/086901}}.

\bibitem{simonsproj}
N.~{Galitzki}, A.~{Ali}, K.~S. {Arnold}, P.~C. {Ashton}, J.~E. {Austermann},
  C.~{Baccigalupi}, et~al.
\newblock {The Simons Observatory: instrument overview}.
\newblock In {\em Millimeter, Submillimeter, and Far-Infrared Detectors and
  Instrumentation for Astronomy IX}, volume 10708 of {\em Society of
  Photo-Optical Instrumentation Engineers (SPIE) Conference Series}, page
  1070804, July 2018.
\newblock \href {http://arxiv.org/abs/1808.04493} {\path{arXiv:1808.04493}},
  \href {http://dx.doi.org/10.1117/12.2312985} {\path{doi:10.1117/12.2312985}}.

\bibitem{CCATP}
G.~J. {Stacey}, M.~{Aravena}, K.~{Basu}, N.~{Battaglia}, B.~{Beringue},
  F.~{Bertoldi}, et~al.
\newblock {CCAT-Prime: science with an ultra-widefield submillimeter
  observatory on Cerro Chajnantor}.
\newblock In {\em Ground-based and Airborne Telescopes VII}, volume 10700 of
  {\em Society of Photo-Optical Instrumentation Engineers (SPIE) Conference
  Series}, page 107001M, July 2018.
\newblock \href {http://arxiv.org/abs/1807.04354} {\path{arXiv:1807.04354}},
  \href {http://dx.doi.org/10.1117/12.2314031} {\path{doi:10.1117/12.2314031}}.

\bibitem{atlast}
Frank Bertoldi.
\newblock {The Atacama Large Aperture Submm/mm Telescope (AtLAST) Project},
  January 2018.
\newblock URL: \url{https://doi.org/10.5281/zenodo.1158842}, \href
  {http://dx.doi.org/10.5281/zenodo.1158842}
  {\path{doi:10.5281/zenodo.1158842}}.

\bibitem{Hargrave2018}
Peter Hargrave.
\newblock Atlast telescope design working group report, January 2018.
\newblock URL: \url{https://doi.org/10.5281/zenodo.1159025}, \href
  {http://dx.doi.org/10.5281/zenodo.1159025}
  {\path{doi:10.5281/zenodo.1159025}}.

\bibitem{Kawabe2016}
R.~{Kawabe}, K.~{Kohno}, Y.~{Tamura}, T.~{Takekoshi}, T.~{Oshima}, and
  S.~{Ishii}.
\newblock {New 50-m-class single-dish telescope: Large Submillimeter Telescope
  (LST)}.
\newblock In {\em Ground-based and Airborne Telescopes VI}, volume 9906 of {\em
  \procspie}, page 990626, August 2016.
\newblock \href {http://dx.doi.org/10.1117/12.2232202}
  {\path{doi:10.1117/12.2232202}}.

\bibitem{Padin2014}
S.~{Padin}.
\newblock {Inexpensive mount for a large millimeter-wavelength telescope}.
\newblock {\em \ao}, 53:4431--4439, July 2014.
\newblock \href {http://dx.doi.org/10.1364/AO.53.004431}
  {\path{doi:10.1364/AO.53.004431}}.

\bibitem{Golwala2018}
S.~{Golwala}.
\newblock {The Chajnantor Sub/Millimeter Survey Telescope}.
\newblock In {\em Atacama Large-Aperture Submm/mm Telescope (AtLAST)}, page~46,
  January 2018.
\newblock \href {http://dx.doi.org/10.5281/zenodo.1159094}
  {\path{doi:10.5281/zenodo.1159094}}.

\bibitem{NIKA2}
R.~{Adam}, A.~{Adane}, P.~A.~R. {Ade}, P.~{Andr{\'e}}, A.~{Andrianasolo},
  H.~{Aussel}, et~al.
\newblock {The NIKA2 large-field-of-view millimetre continuum camera for the 30
  m IRAM telescope}.
\newblock {\em \aap}, 609:A115, January 2018.
\newblock \href {http://arxiv.org/abs/1707.00908} {\path{arXiv:1707.00908}},
  \href {http://dx.doi.org/10.1051/0004-6361/201731503}
  {\path{doi:10.1051/0004-6361/201731503}}.

\bibitem{Molnar2009}
Sandor~M. {Molnar}, Nathan {Hearn}, Zolt{\'a}n {Haiman}, Greg {Bryan},
  August~E. {Evrard}, and George {Lake}.
\newblock {Accretion Shocks in Clusters of Galaxies and Their SZ Signature from
  Cosmological Simulations}.
\newblock {\em \apj}, 696:1640--1656, May 2009.
\newblock \href {http://arxiv.org/abs/0902.3323} {\path{arXiv:0902.3323}},
  \href {http://dx.doi.org/10.1088/0004-637X/696/2/1640}
  {\path{doi:10.1088/0004-637X/696/2/1640}}.

\bibitem{Kitayama2016}
Tetsu {Kitayama}, Shutaro {Ueda}, Shigehisa {Takakuwa}, Takahiro {Tsutsumi},
  Eiichiro {Komatsu}, Takuya {Akahori}, et~al.
\newblock {The Sunyaev-Zel'dovich effect at 5″: RX J1347.5-1145 imaged by
  ALMA}.
\newblock {\em Publications of the Astronomical Society of Japan}, 68:88, Oct
  2016.
\newblock \href {http://arxiv.org/abs/1607.08833} {\path{arXiv:1607.08833}},
  \href {http://dx.doi.org/10.1093/pasj/psw082}
  {\path{doi:10.1093/pasj/psw082}}.

\end{thebibliography}

\end{document}